\documentclass[letterpaper, 12pt, titlepage, reqno]{article}

\usepackage[letterpaper, portrait, margin=2.5cm]{geometry}

 \usepackage[super]{natbib}
 \usepackage{amsmath, amssymb, latexsym}

\usepackage{graphicx}% Include figure files
\usepackage{epsfig}
\usepackage{epstopdf}
\usepackage{caption}

\renewcommand{\epsilon}{\varepsilon}

\newcommand{\bc}{\mbox{${\bf c}$}}

\newcommand{\bu}{\mbox{${\bf u}$}}

\newcommand{\bepsilon}{\mbox{\boldmath $\bf \varepsilon$}}

\newcommand{\bsigma}{\mbox{\boldmath $\bf \sigma$}}

\newcommand{\bdel}{\mbox{\boldmath $\nabla$}}

\DeclareMathOperator{\Tr}{Tr}

\begin{document}

 \begin{titlepage}
 \begin{center}

\textbf{An axisymmetric time-domain spectral-element method for full-wave simulations: Application to ocean acoustics}\\

 \vspace{10ex}

Alexis Bottero, Paul Cristini and Dimitri Komatitsch\footnote{e-mail: komatitsch@lma.cnrs-mrs.fr}\\
LMA, CNRS UPR 7051, Aix-Marseille Univ., Centrale Marseille,\\
13453 Marseille cedex 13, France.\\

 \vspace{5ex}

Mark Asch\\
LAMFA, CNRS UMR 7352, Universit\'e de Picardie Jules Verne, Amiens, France.
 \end{center}
 \end{titlepage}

\begin{abstract}
The numerical simulation of acoustic waves in complex 3D media is a key topic in many branches of science, from exploration geophysics to non-destructive testing and medical imaging. With the drastic increase in computing capabilities this field has dramatically grown in the last twenty years. However many 3D computations, especially at high frequency and/or long range, are still far beyond current reach and force researchers to resort to approximations, for example by working in 2D (plane strain) or by using a paraxial approximation. This article presents and validates a numerical technique based on an axisymmetric formulation of a spectral finite-element method in the time domain for heterogeneous fluid-solid media. Taking advantage of axisymmetry enables the study of relevant 3D configurations at a very moderate computational cost. The axisymmetric spectral-element formulation is first introduced, and validation tests are then performed. A typical application of interest in ocean acoustics showing upslope propagation above a dipping viscoelastic ocean bottom is then presented. The method correctly models backscattered waves and explains the transmission losses discrepancies pointed out in Jensen et al. (2007). Finally, a realistic application to a double seamount problem is considered.
\end{abstract}

\noindent \begin{center}{\textbf{*** This manuscript is now published as a paper in the Journal of the Acoustical Society of America, 2017. ***}}\end{center}

\section{Introduction}\label{introduction}

Among all the numerical methods that can be used to model acoustic wave propagation in the ocean (see e.g. \citet{COA2011} for a
comprehensive review), finite-element techniques are methods of choice for solving the wave or Helmholtz equation very accurately and for handling
complex geometries\cite{KoTr99,Fic10,COA2011}.
They have for instance been used to study acoustic scattering by rough interfaces\cite{IsCh11,IsCh15}, which is a topic of importance in reverberation studies
and ocean bottom sensing, and for the study of diffraction by structures immersed or embedded within the oceanic medium\cite{ZaTeJeMaBl07}.
When time signals are needed they can be obtained by Fourier synthesis from frequency-based simulations, or else obtained directly by solving the wave equation
in the time domain.
In this case, spectral-element methods\cite{CrKo12,JaGuGuMaRo13} are of interest for performing numerical simulations
because they lead to accurate results and an efficient implementation\cite{KoTr99,Fic10,Kom11,PeKoLuMaLeCaLeMaLiBlNiBaTr11}.
In addition they are very well adapted to modern computing clusters\cite{TrKoHjLiZhPeBoMcFrTrHu10} and supercomputers.
However, finite-element models have the drawback of requiring large computational resources compared to approximate numerical methods
that do not solve the full-wave equation, for instance the parabolic approximation\cite{PiTr08}.
Despite the recent development of very efficient perfectly matched absorbing layers for the study of wave propagation in fluid-solid regions\cite{XiMaCrKoMa16},
which allows for a drastic reduction of the size of the computational domain, performing realistic 3D simulations in ocean acoustics still turns out to be
difficult
because of the high computational cost incurred. The main reason for this lies in the size of the domains classically studied in ocean acoustics
that often represent thousands of wavelengths.
A first attempt was presented at low frequency in \citet{XiMaCrKoMa16},
but when higher frequencies are required in the context of full-wave simulations,
2D simulations are currently still the only option. In previous work\cite{CrKo12} on underwater acoustics, we used a 2D Cartesian (plane strain) version of the
spectral-element method.
This type of 2D simulation has the disadvantage of involving line sources, i.e. unphysical sources that extend in the direction
perpendicular to the 2D plane\cite{Pil79}. Physical effects are thus enhanced in an artificial manner
and may lead to erroneous interpretation. For the same reason, comparisons with real data are difficult because amplitudes and waveforms are unrealistic.

A more realistic approach thus consists of resorting to axisymmetric simulations.
If the source is situated on the symmetry axis then this allows for the calculation of wavefields having 3D geometrical spreading with the cost of a 2D simulation. Nevertheless,
solving the weak form of the wave equation in cylindrical coordinates leads to a difficulty because of the potential singularity in elements that are in contact
with the symmetry axis, for which $r = 0$ at some of their points and thus the factor $1/r$ becomes singular.
Several strategies have been proposed to overcome this difficulty. One approach consists of increasing the accuracy of
the numerical integration when an element is close to the axis or lies on the axis and slightly shifting its position away from the axis\cite{ClRe00}.
However, in such a case the source cannot be put on the axis, making simulations of point sources impossible.
A better strategy was developed by \citet{BeDaMaAz99}, who proposed to use a Gauss-Radau-type quadrature rule to handle
this problem and remove the singularity. This approach was successful and several subsequent articles have implemented axisymmetric simulations using
this strategy to reduce the computational cost of spectral-element-based numerical simulations. For instance, \citet{FoBuHoVi04} used axisymmetric
cylindrical coordinates to simulate thermal convection in fluid-filled containers,
and Nissen-Meyer et al. \cite{NiFoDa07,NiFoDa08,NiVaStHoHeAuCoFo14} used a clever combination of four runs performed in cylindrical coordinates to
simplify the computations of seismic wavefields in spherical coordinates in both fluid-solid and purely solid regions.
However, to our knowledge ocean acoustics configurations with axisymmetric
cylindrical coordinates have never been addressed. In this article we thus present an axisymmetric spectral-element method in cylindrical coordinates
for such fluid-solid models that is particularly well-adapted to underwater acoustics.

The article is organized as follows: In Section II we describe the implementation of an axisymmetric spectral-element method in cylindrical coordinates for
fluid and solid regions. Section III is then devoted to the validation of this implementation by providing comparisons with results obtained with analytical or
other numerical methods. We first consider a flat bottom configuration, and then analyze a slope bottom configuration that exhibits
significant backscattering, illustrating the interest of performing full-wave time-domain simulations.
Finally, in Section IV we illustrate the importance of taking backscattering into account in acoustic wave propagation in the ocean
based on a configuration with two seamounts.
This model exhibits strong backscattering effects that can only be modeled based on a full-wave simulation.

%%%%%%%
\section{Axisymmetric spectral elements}
%%%%%%%

In this section we will briefly present our axisymmetric spectral-element technique.
Interested readers  can learn more on spectral element methods in their Cartesian forms
by referring to \citet{Coh02,DeFiMu02} or \citet{Fic10} for example.
Let $\boldsymbol{x}$ denote the position vector.
In the general case, the time-dependent displacement field $\boldsymbol{u}(\boldsymbol{x},t)$ induced by an acoustic source
$\boldsymbol{f}(\boldsymbol{x},t)$ is related to the medium features by the 3D wave equation, which can be written in its strong form as
\begin{equation}
\rho\ddot{\boldsymbol{u}}=\boldsymbol{\nabla}\cdot\boldsymbol{\sigma}+\boldsymbol{f},\label{eq:momentum}
\end{equation}
where $\rho(\boldsymbol{x})$ is the distribution of density, $\boldsymbol{\nabla}\cdot\boldsymbol{\sigma}$ is the divergence of
the stress tensor $\boldsymbol{\sigma}(\boldsymbol{x},t)$,
and a dot over a symbol denotes time differentiation.
In a typical forward problem the source $\boldsymbol{f}(\boldsymbol{x},t)$ and the material properties of the medium are known and we are interested
in computing the displacement field $\boldsymbol{u}(\boldsymbol{x},t)$.

\begin{figure}
\centerline{\includegraphics[width=0.8\linewidth]{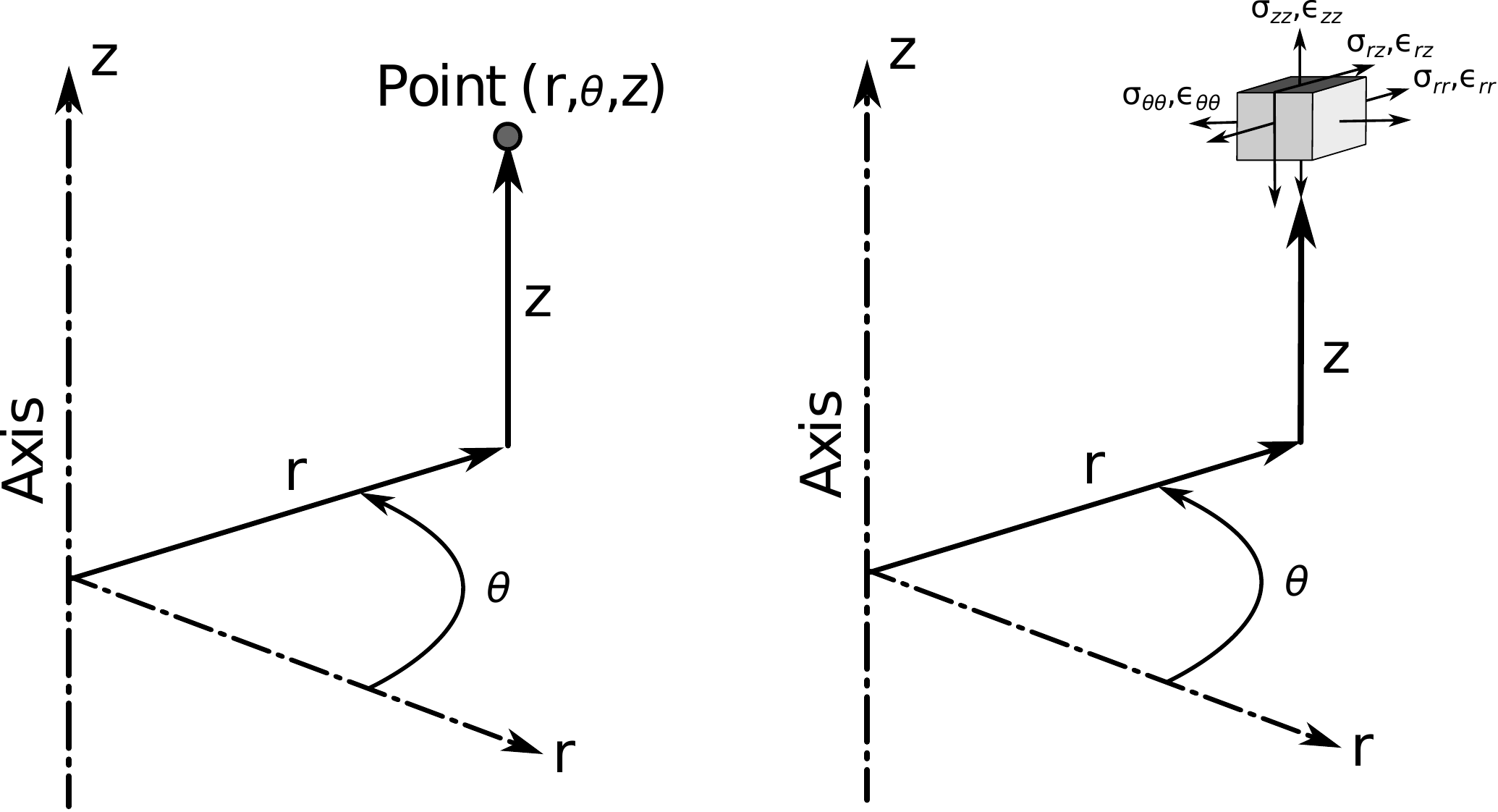}}
\caption{\label{fig:cyl} Cylindrical coordinate system and components of the strain tensor $\boldsymbol{\varepsilon}$
and of the stress tensor $\boldsymbol{\sigma}$ in an axisymmetric setting.}
\end{figure}

\begin{figure}
\centerline{\includegraphics[width=0.495\linewidth]{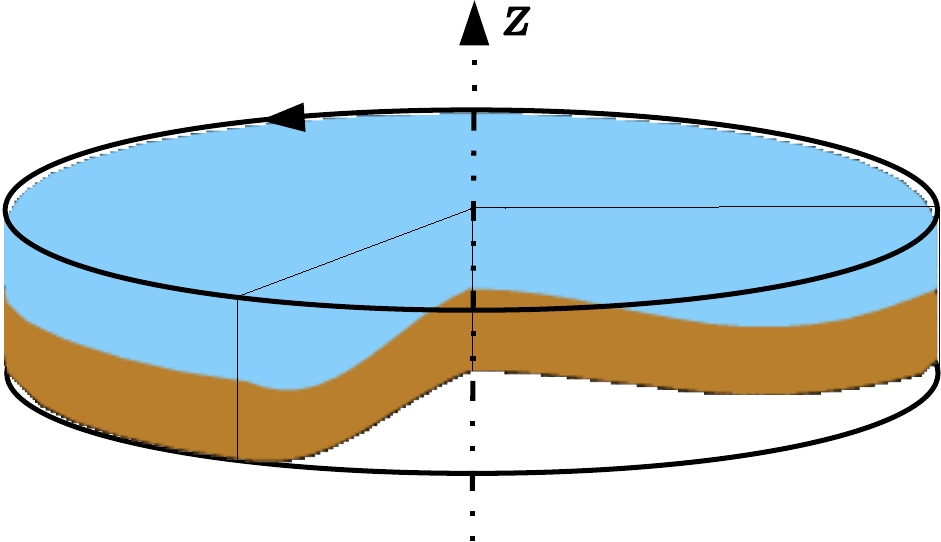}\hfill\includegraphics[width=0.495\linewidth]{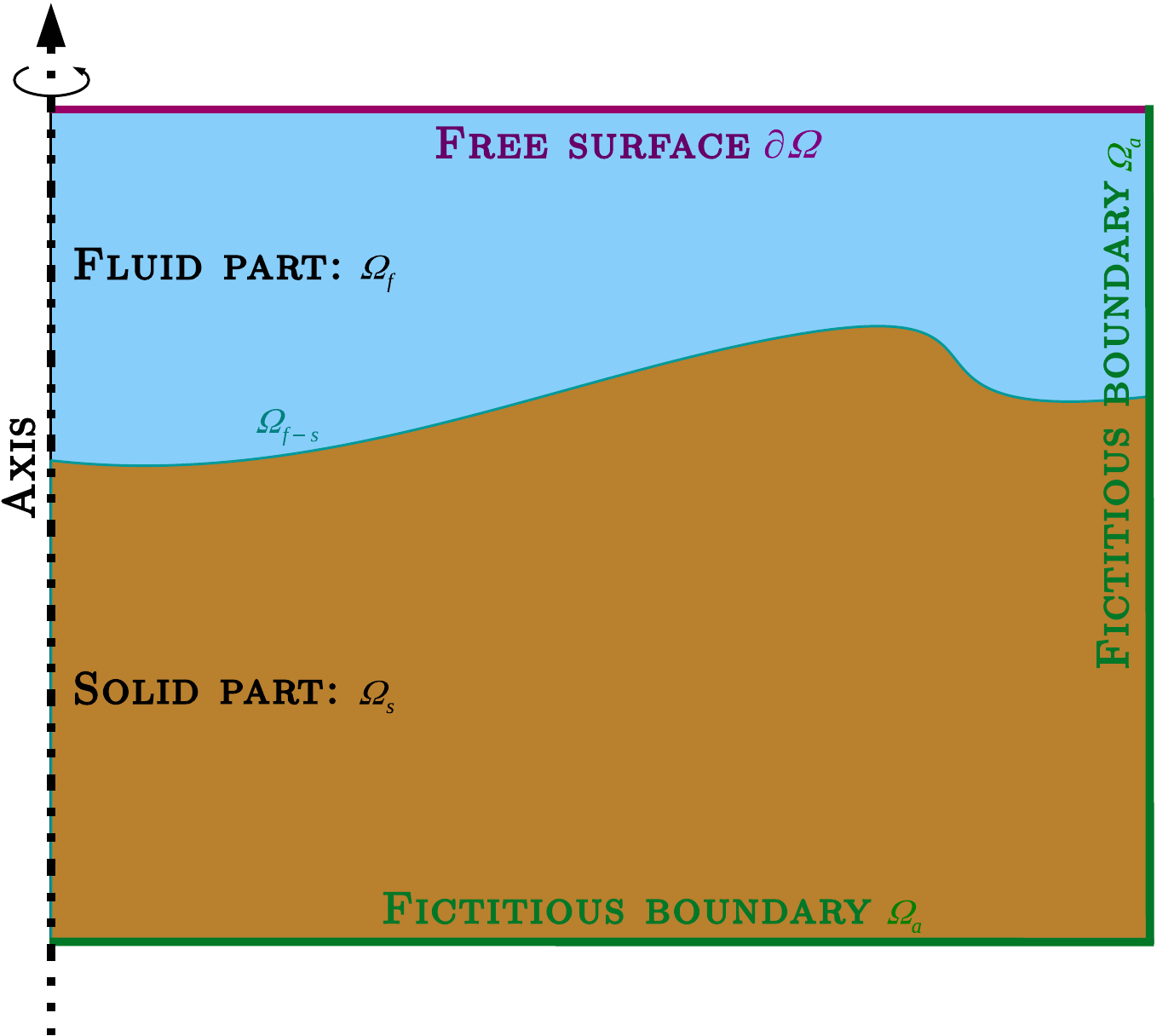}}
\caption{Left: Axisymmetric medium generated by rotation of its 2D meridional shape around the $z$-axis.
Right: Meridional 2D shape $\Omega$ and description of our notation.}
\label{fig:notations}
\end{figure}

From now on we will choose the cylindrical coordinate system $(r,\theta,z)$ - see Figure \ref{fig:cyl}.
The position vector is then expressed as $\boldsymbol{x}=r\hat{\boldsymbol{r}}+\theta\hat{\boldsymbol{\theta}}+z\hat{\boldsymbol{k}}$ and
any vector $\boldsymbol{a}$ is decomposed into its cylindrical components,
\begin{equation}
\boldsymbol{a}=a_{r}(r,\theta,z,t)\hat{\boldsymbol{r}}+a_{\theta}(r,\theta,z,t)\hat{\boldsymbol{\theta}}+a_{z}(r,\theta,z,t)\hat{\boldsymbol{k}}.
\end{equation}
At this point we choose the 2.5-D convention: we consider a 3D domain $\breve{\Omega}$ symmetric with respect to the axis $(r=0)$ and suppose that the
important loads are axisymmetric also and are thus independent of $\theta$. Then, all the quantities of interest are independent of $\theta$  and, due to the
rotational symmetry, the $\theta$-component of the displacement is zero.
This yields true axisymmetry resulting in a reduction of the order and the number of equations,
while still preserving the possibility of non-zero out-of-plane components of stress, $\sigma_{\theta \theta}$.
An example of such a configuration can be seen in Figure \ref{fig:notations} (left).

Following Curie's symmetry principle, the symmetry of a cause is always preserved in its effects. Hence if the acoustic source
$\boldsymbol{f}(r,\theta,z,t)=f_r(r,z,t)\hat{\boldsymbol{r}}+f_\theta(r,z,t)\hat{\boldsymbol{\theta}}+f_z(r,z,t)\hat{\boldsymbol{k}}$
does not depend on $\theta$, working in the 3D domain $\breve{\Omega}$ reduces to working in its meridional 2D shape $\Omega$,
referred to as the transect in underwater acoustics.
As shown in Figure \ref{fig:notations} (right), in the general case, this 2D meridional shape
$\Omega$ is composed of a fluid part $\Omega_f$ and a solid part $\Omega_s$.
The coupling interface between these two sub-domains is denoted $\Omega_{f-s}$. Let us suppose that $\Omega$ has a free surface $\partial\Omega$,
and fictitious absorbing boundaries $\Omega_{a}$. These are required because for regional simulations our modeling domain is limited by fictitious
boundaries beyond which we are not interested in the wave field and thus acoustic energy reaching the edges of the model needs to be absorbed by our algorithm.
In recent years, an efficient absorbing condition
called the Perfectly Matched Layer (PML) has been introduced\cite{Ber94} and is now used widely in regional numerical simulations.
Although we have implemented them in our code, the mathematical and numerical complications associated with absorbing boundary conditions are beyond
the scope of this article and will not be addressed here as their implementation in the axisymmetric case is very similar to the Cartesian one. The
reader is referred to \citet{MaKoGe08}, \citet{Mat11} or \citet{XiKoMaMa14} for example. Let us just mention that grazing incidence
problems have been solved\cite{XiKoMaMa14} and these enhanced PML layers are implemented in our code.
This allows us to perform simulations on very elongated domains.
For the sake of completeness we will note $\Gamma=\partial\Omega \, \cup \, \Omega_{f-s} \, \cup \, \Omega_{a}$
all the possible one-dimensional boundaries considered,
although we will focus on fluid-solid interfaces $\Omega_{f-s}$.
For $\boldsymbol{x} \in \Gamma$ we note $\boldsymbol{n}(\boldsymbol{x})$ the outward unit normal to the surfaces $\Gamma$.
The spectral-element method, just as the standard finite-element method, is based on a weak form of the wave equation \eqref{eq:momentum}.
This formulation being different in the fluid and solid parts of the model we will present each case separately.
It is worth mentioning that unified fluid-solid formulations can be designed if needed\cite{WiStBuGh10},
however they lead to a larger number of calculations on the fluid side.

\subsection{Fluid parts}

\begin{figure}
\centerline{\includegraphics[width=0.45\linewidth]{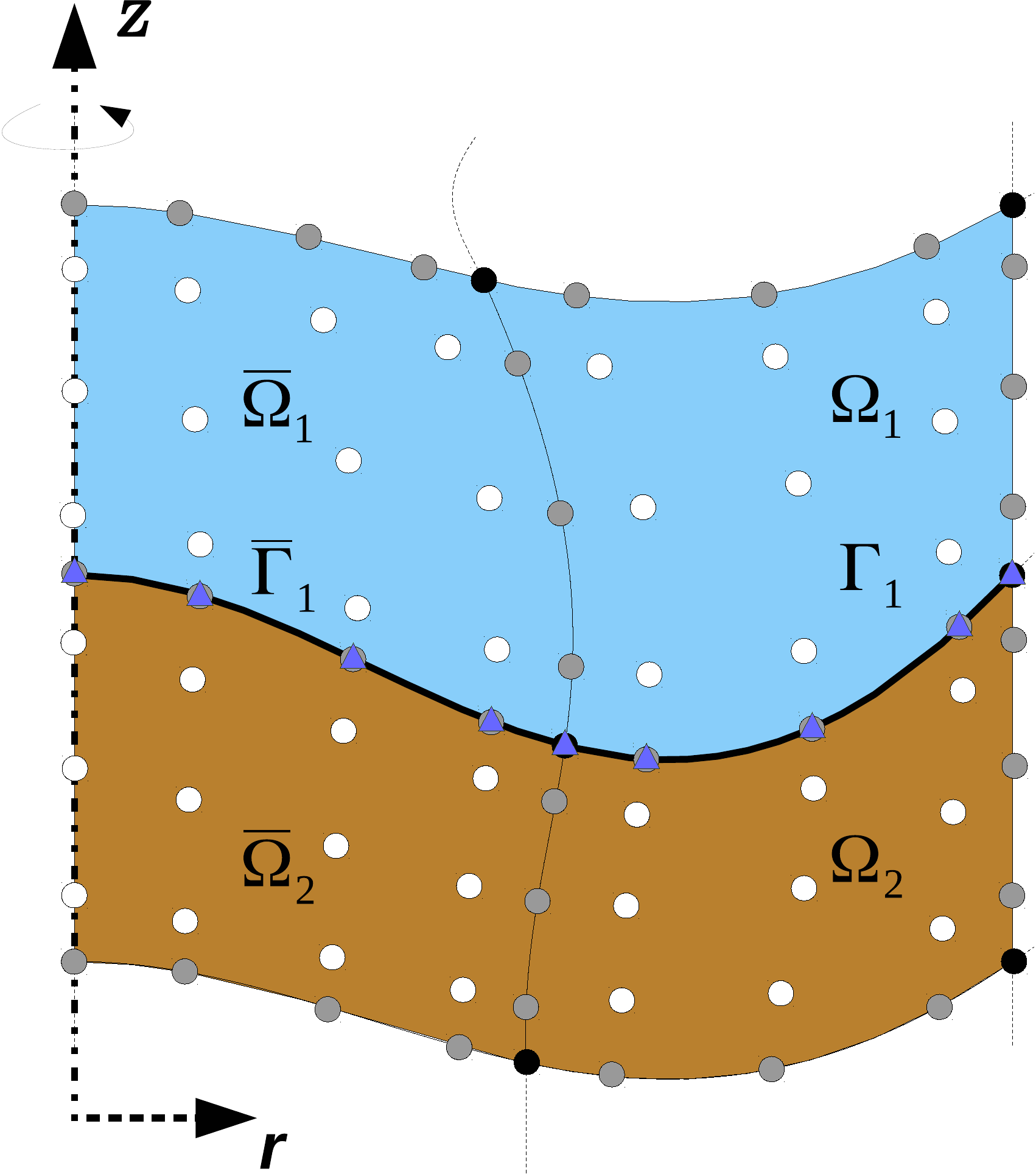}}
\caption{\label{fig:GLpoints} Typical setup along the symmetry axis for a 2D model containing two media.
Four 2D elements of polynomial degree $N=4$ are shown. The two elements at the top
($\overline{\Omega}_1$ and $\Omega_1$) lie in a fluid medium, the two at the bottom ($\overline{\Omega}_2$ and $\Omega_2$) lie in an elastic medium,
and two 1D edges ($\overline{\Gamma}_1$ and $\Gamma_1$) are used to describe the coupling along the interface.
The circles represent the GLL points, or the GLJ points along the direction $r$ in the case of the elements that are in contact with the axis.
The triangles are the GLL/GLJ points on the 1D edges.
Each 2D spectral element contains $(N+1)^2 = 25$ GLL (or GLJ) points, which constitute the local mesh for each element.
The difference between local and global grids appears here: points lying on edges or corners are shared amongst several elements,
and thus the contributions to the global system, computed separately for each element, have to be summed (i.e. assembled)
at these common points represented by black or gray circles.}
\end{figure}

\begin{figure}
\centerline{\includegraphics[width=0.38\linewidth]{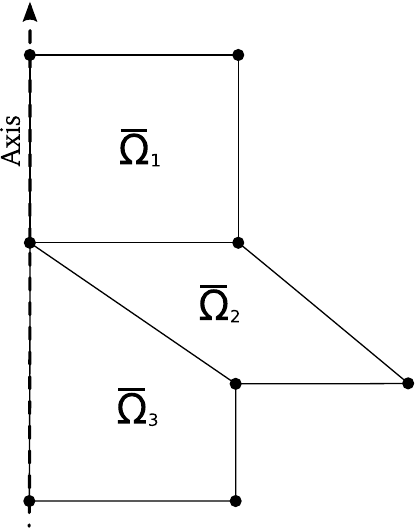}}
\caption{\label{fig:meshrestrictionontheaxis} For simplicity we exclude cases in which the mesh elements that are in contact with the symmetry axis
are in contact with it by a single point instead of by a full edge, such as element $\bar{\Omega}_2$ here.
This amounts to imposing that the leftmost layer of elements in the mesh be structured rather than non structured; The rest of the mesh can be non structured.}
\end{figure}

In the fluid regions, and ignoring the source term for now, the wave equation \eqref{eq:momentum} when written for pressure $P(\boldsymbol{x},t)$ in a
spatially-heterogeneous fluid is\cite{BreGod90}
\begin{equation}
\dfrac{1}{\kappa}\ddot{P}=\boldsymbol{\nabla}\cdot\left(\dfrac{\boldsymbol{\nabla}P}{\rho}\right) \, ,
\label{eqforpressure}
\end{equation}
where $\kappa(\boldsymbol{x})$ is the adiabatic bulk modulus of the fluid.
The linearized Euler equation is valid in a fluid with constant or spatially slowly-varying density\cite{LaLi59,COA2011} and reads
\begin{equation}
\ddot{\boldsymbol{u}} = - \dfrac{\boldsymbol{\nabla}P}{\rho} \, .
\label{eulereq}
\end{equation}
Following \citet{Eve81} and then \citet{ChVa04}, from a numerical point of view it is more convenient to integrate this system twice in time because that
allows for numerical fluid-solid coupling based on a non-iterative scheme. We thus define a new scalar potential,
\begin{equation}
\ddot{\chi}=-P
\label{definepotential}
\end{equation}
and the scalar wave equation (\ref{eqforpressure}) then becomes
\begin{equation}
\dfrac{1}{\kappa}\ddot{\chi}=\boldsymbol{\nabla}\cdot\left(\dfrac{\boldsymbol{\nabla}\chi}{\rho}\right) \, .
\end{equation}
Equation (\ref{eulereq}) combined with (\ref{definepotential}) leads to
\begin{equation}
\rho \boldsymbol{u} = \boldsymbol{\nabla}\chi \, ,
\end{equation}
i.e. $\rho \boldsymbol{u}$ is irrotational\cite{COA2011}.
Adding a pressure point-source at position $\boldsymbol{x}_s$ we obtain the wave equation in fluids,
\begin{equation}
\dfrac{1}{\kappa}\ddot{\chi}=\boldsymbol{\nabla}\cdot\left(\dfrac{\boldsymbol{\nabla}\chi}{\rho}\right) + \dfrac{1}{\kappa}f(t)\delta_{\boldsymbol{x}_s} \, .
\label{eq:wavesFluids}
\end{equation}
Let $\boldsymbol{x}\mapsto w(\boldsymbol{x})$ be a real-valued, arbitrary test function defined on $\Omega_f$.
One obtains the weak form by dotting the wave equation \eqref{eq:wavesFluids} with a scalar test function $w$
and integrating by parts over the model volume $\Omega_f$,
\begin{equation}
\begin{array}{ccl}
{\displaystyle \int_{\Omega_f}w\dfrac{1}{\kappa}\ddot{\chi}\:\mathrm{d^{2}\boldsymbol{x}}} & = &
{\displaystyle-\int_{\Omega_f}\dfrac{1}{\rho}\boldsymbol{\nabla}w\cdot\boldsymbol{\nabla}\chi\:\mathrm{d^{2}\boldsymbol{x}}
+ \int_{\Omega_{f-s}}\dfrac{1}{\rho}w\boldsymbol{n}\cdot\dot{\boldsymbol{u}}\:\mathrm{d\Gamma}
+ \dfrac{1}{\kappa(\boldsymbol{x}_s)}w(\boldsymbol{x}_s) f(t) \, .
}
\end{array}
\label{eq:weakForm3Dacoust}
\end{equation}
Note that the infinitesimal surface is now $\mathrm{d^{2}\boldsymbol{x}} = 2\pi r\mathrm{dr dz}$.
Remark that the part of the contour integral along the free surface $\partial\Omega$ has vanished. Indeed, the pressure-free surface condition
is $P(\boldsymbol{x},t)=-\ddot{\chi}(\boldsymbol{x},t)=0$ for all times and all $\boldsymbol{x}\in\partial\Omega$,
and thus $\dot{\chi}(\boldsymbol{x},t)=0$ as well, hence $\boldsymbol{n}\cdot\dot{\boldsymbol{u}}=\boldsymbol{n}\cdot(\frac{1}{\rho}\nabla\dot{\chi})=0$,
making the contour integral along the free surface $\partial\Omega$ vanish naturally.
The first term of \eqref{eq:weakForm3Dacoust} is traditionally called the mass integral,
the second is the stiffness integral and the third is the fluid-solid coupling integral.
The last term is the (known) source term, and will be dropped for notational convenience.
It should be noted that both the test function $w(r,z)$ and the radial derivative of the potential $\frac{\partial \chi}{\partial r}(r,z,t)$
have to vanish on the axis\cite{BeDaMaAz99}.
Let us now present the discretization of this equation based on the time-domain spectral-element method. This is somewhat similar
to the 2D planar spectral-element implementation (of e.g. \citet{CrKo12}),
thus for the sake of conciseness we detail here the differences only.
We recall that the model $\Omega$ is subdivided into a number of non-overlapping quadrangular elements $\Omega_{e}$ , $e=1,\ldots,n_e$,
such that $\Omega=\bigcup_{e=1}^{n_e}\Omega_{e}$. As a result of this subdivision, the boundary $\Gamma$ is similarly represented by a number
of 1D edges $\Gamma_{b}$, $b=1,\ldots,n_b$, such that $\Gamma=\bigcup_{b=1}^{n_b}\Gamma_{b}$.
The $\bar{n}_e$ 2D elements that are in contact with the axis need to be distinguished, they will be noted $\overline{\Omega}_e$.
Similarly we will note $\overline{\Gamma}_b$ the 1D horizontal edges that are in contact with the axis by one point (Figure~\ref{fig:GLpoints}).
For simplicity we also assume that the mesh elements that are in contact with the symmetry axis are in contact with it
by a full edge rather than by a single point, i.e. we exclude cases such as that of Figure~\ref{fig:meshrestrictionontheaxis}.
This amounts to imposing that the leftmost layer of elements in the mesh be structured rather than non structured, while the rest of the mesh can be non structured.

The integrals in the weak form (\ref{eq:weakForm3Dacoust}) are then split into integrals over the elements, in turn expressed as integrals
over 2D and 1D reference elements $\Lambda=[-1,1]\times[-1,1]$ and $[-1,1]$ thanks to an invertible mapping between global coordinates $(r,z)$
and reference local coordinates $(\xi,\eta)$:
\begin{equation}
\begin{array}{ccl}
{\displaystyle \int_{\Omega_{e}\text{ or }\bar{\Omega}_{e}}w\dfrac{1}{\kappa}\ddot{\chi}\:r\mathrm{dr}\mathrm{dz}} & = & {\displaystyle
\int_{\Lambda}w\dfrac{1}{\kappa}\ddot{\chi}\:\left|\mathcal{J}_{e}\right|r\mathrm{d\xi}\mathrm{d\eta}} , \\
{\displaystyle \int_{\Omega_{e}\text{ or }\bar{\Omega}_{e}}\dfrac{1}{\rho}\boldsymbol{\nabla}w\cdot\boldsymbol{\nabla}\chi\: r\mathrm{dr}\mathrm{dz}} & = &
{\displaystyle \int_{\Lambda}\dfrac{1}{\rho}\boldsymbol{\nabla}w\cdot\boldsymbol{\nabla}\chi\:\left|\mathcal{J}_{e}\right|r\mathrm{d\xi}\mathrm{d\eta}}  , \\
{\displaystyle \int_{\Gamma_{e}\text{ or }\bar{\Gamma}_{e}}w\boldsymbol{n}\cdot\dot{\boldsymbol{u}}\: r\mathrm{d\Gamma}} & = & {\displaystyle
\int_{-1}^{1}w\boldsymbol{n}\cdot\dot{\boldsymbol{u}}\:\dfrac{dr}{d\xi}r\mathrm{d\xi}} ,
\end{array}
\end{equation}
where $\left|\mathcal{J}_{e}\right|$ is the Jacobian of the invertible mapping. It then remains to calculate these integrals.
The spectral-element method is based upon a high-order piecewise polynomial approximation of the weak form of the wave equation.
For non axial elements, integrals along $\xi$ and $\eta$ are computed based upon Gauss-Lobatto-Legendre (GLL) quadrature.
The integral of a function is expressed as a weighted sum of the values of the function at $N$ specified collocation points called GLL points
(containing $-1$ and $1$) as described in textbooks on the spectral-element method\cite{Coh02,DeFiMu02,Fic10}.
The $N$ GLL points along $\xi$ will be noted $\xi_i$, and those along $\eta$ are noted $\eta_i$, the integration weights associated are noted $\omega_i$ and
the basis functions are $\xi \mapsto \ell_i(\xi)$ or $\eta \mapsto \ell_i(\eta)$.
In the axisymmetric case, following \citet{BeDaMaAz99},\citet{GePh00},\citet{FoBuHoVi04},\citet{NiFoDa07} and \citet{NiFoDa08} we use a different quadrature
for the integration in the $\xi$-direction for elements that are in contact with the symmetry axis.
Indeed one can see that the factor $r$ in the infinitesimal surface $r\mathrm{d\xi d\eta}$ would lead to undetermined equations, $0 = 0$,
if the integrands were to be evaluated at $\xi_0=-1$.
To deal with this issue the easiest solution would be to shift the edge of the mesh by a small distance
away from the axis (see e.g. \citet{KaDo99} or \citet{ClRe00})
but this convenient solution is not very meaningful for most physical problems because the source, not being on the axis, has to have a circular shape
and thus an unphysical radiation pattern.
\citet{BeDaMaAz99} introduced a convenient way to tackle this issue, which has now become the classical approach
for axisymmetric spectral-element problems\cite{GePh00,FoBuHoVi04,NiFoDa07,NiFoDa08}.
In the $\eta$ direction (the vertical direction) nothing different has to be done,
but in the $\xi$ direction one resorts to Gauss-Lobatto-Jacobi (GLJ) quadrature $(0,1)$.
One first defines the set of polynomials $\overline{P}_{N}$ based on $N^{\text{th}}$-degree Legendre polynomials $P_N$ following the relation,
\begin{equation}
\overline{P}_{N}(\xi)=\dfrac{P_{N}(\xi)+P_{N+1}(\xi)}{1+\xi}.
\end{equation}
The $N+1$ GLJ points $\overline{\xi}_{i}$ are then the $N+1$ zeros of $\xi \mapsto (1-\xi^{2})\dfrac{d\overline{P}_{N}}{d\xi}(\xi)$,
and one computes the GLJ basis functions\cite{NiFoDa07} $\xi \mapsto \overline{\ell}_{i}(\xi)$,
\begin{equation}
\overline{\ell}_{i}(\xi)=\begin{cases}
\dfrac{2(-1)^{N}(\xi-1)\dfrac{\partial\overline{P}_{N}(\xi)}{\partial\xi}}{N(N+1)(N+2)}, & i=0 ,\\
\dfrac{1}{N(N+2)\overline{P}_{N}(\overline{\xi}_{i})}\dfrac{(1-\xi^{2})\dfrac{\partial\overline{P}_{N}(\xi)}{\partial\xi}}{\overline{\xi}_{i}-\xi}, & 0<i<N ,\\
\dfrac{(1+\xi)\dfrac{\partial\overline{P}_{N}(\xi)}{\partial\xi}}{N(N+2)}, & i=N.
\end{cases}
\end{equation}
Any function $g:(r,z)\mapsto g(r,z)$ on $\overline{\Omega}_{e}$ can be decomposed on its values at the GLJ and GLL points,
\begin{equation}
\forall \boldsymbol{\xi}=(\xi,\eta)\in\Lambda\quad
g(\boldsymbol{x}(\boldsymbol{\xi}))
\approx \sum_{\alpha=0}^{N}\sum_{\beta=0}^{N}g(\boldsymbol{x}(\overline{\xi}_{\alpha},\eta_{\beta}))\overline{\ell}_{\alpha}(\xi)\ell_{\beta}(\eta)
\equiv \sum_{\alpha,\beta=0}^{N}g^{\overline{\alpha}\beta}\overline{\ell}_{\alpha}(\xi)\ell_{\beta}(\eta),
\end{equation}
where we have noted $g^{\overline{\alpha}\beta}\equiv g(\boldsymbol{x}(\overline{\xi}_{\alpha},\eta_{\beta}))$ the value of $g$ at the GLJ/GLL points
$(\overline{\xi}_{\alpha},\eta_{\beta})$. It is worth mentioning the important property $\overline{\ell}_{i}(\overline{\xi}_{j})=\ell_{i}(\xi_{j})=\delta_{ij}$
as well as the fact that the points $-1$ and $1$ still belong to the set of collocation points: the continuity with non-axial elements is thus ensured,
and the mass matrix remains diagonal.
We evaluate the surface integrals with the following quadrature rule,
\begin{equation}
\int_{\Lambda}g\:\mathrm{d\xi}\mathrm{d\eta}\approx\sum_{i,j=0}^{N}\overline{\omega}_{i}\omega_{j}\dfrac{g^{\overline{i}j}}{\overline{\xi}_{i}+1},
\end{equation}
where the $\overline{\omega}_{k}=\int_{-1}^1\overline{\ell}_{k}(\xi)\:\mathrm{d\xi}$ are the GLJ integration weights.
As the GLJ points, they are computed easily once and for all. All functions $g$ considered in axial elements satisfy $g^{\overline{0}\beta}=0$,
consequently the singularities involving the terms $\frac{g^{\overline{i}j}}{\overline{\xi}_{i}+1}$ are easily removed with l'H\^opital's rule\cite{BeDaMaAz99},
\begin{equation}
\underset{\xi\rightarrow\overline{\xi}_{0}}{\lim}\dfrac{g(\boldsymbol{x}(\overline{\xi},\eta))}{\xi+1}
=\frac{\partial g}{\partial\xi}(\boldsymbol{x}(\overline{\xi}_{0},\eta))
={\displaystyle \sum_{\alpha,\beta=0}^{N}g^{\overline{\alpha}\beta}\overline{\ell}'_{\alpha}(\overline{\xi}_{0})\ell_{\beta}(\eta)}.
\end{equation}
The 1D integrals on horizontal edges of axial elements (the only ones that are different from the Cartesian formulation) are computed in the same way,
forgetting the variable $\eta$. For any function $g:(r)\mapsto g(r)$ on $\overline{\Gamma}_{e}$,
\begin{equation}
      {\displaystyle \forall \xi\in [-1,1]\quad g(r(\xi))\approx\sum_{\alpha=0}^{N}g(r(\overline{\xi}_{\alpha}))\overline{\ell}_{\alpha}(\xi)
      \equiv \sum_{\alpha=0}^{N}g^{\overline{\alpha}}\:\overline{\ell}_{\alpha}(\xi)}
\end{equation}
and
\begin{equation}
      \int_{-1}^1 g\:\mathrm{d\xi}\approx\sum_{i=0}^{N}\overline{\omega}_{i}\dfrac{g^{\overline{i}}}{\overline{\xi}_{i}+1}.
\end{equation}
After substituting these representations in the wave equation \eqref{eq:weakForm3Dacoust}, a last step has to be performed to deal with GLL/GLJ points lying on
the sides, edges or corners of an element that are shared amongst neighboring elements (see Figure \ref{fig:GLpoints}).
Therefore, the grid points that define an element (the local mesh) and all the grid points in the model (the global mesh) need to be distinguished and mapped.
Efficient routines are available for this purpose from finite-element modeling. Before the system can be marched forward in time, the contributions
from all the elements that share a common global grid point need to be summed. In traditional finite-element this is referred to as the assembly of the system.
For more details on the matrices involved and on the assembly of the system the reader can refer e.g. to the Appendix of \citet{KoTr99} or to \citet{Fic10}.
Using the fact that the relations must hold for any test function $w$, one obtains a global algebraic system of equations,
\begin{equation}
{\displaystyle \ddot{\boldsymbol{\chi}}}=\left({\boldsymbol{M}}\right)^{-1}\boldsymbol{F}_{int}(t) ,
\label{eq:globalSystem}
\end{equation}
in which $\boldsymbol{\chi}$ gathers the value of the potential and $\boldsymbol{F}_{int}(t)$ gathers all the interior forces at any unique GLL/GLJ point.
The term $\boldsymbol{M}$ is the mass matrix. By construction in the spectral-element method
this matrix is diagonal and its inversion is thus straightforward and does not incur any significant
computational cost. Hence time discretization of the second-order hyperbolic ordinary differential equation \eqref{eq:globalSystem} can be based upon
a fully explicit time scheme.  In practice we select an explicit second-order-accurate finite-difference scheme, which is a particular case of the
Newmark scheme\cite{Hug87}. This scheme is conditionally stable, and the Courant stability condition is governed by the maximum value of the ratio
between the compressional wave speed and the grid spacing. The main numerical cost associated with the spectral-element method is related to small, local
matrix-vector products between the local field and the local stiffness matrix,
not to the time-integration scheme\cite{TrKoLi08}.

\subsection{Solid parts}

In linear elastic solids the strain tensor $\boldsymbol{\varepsilon}(\boldsymbol{x},t)$ is calculated from the displacement vector $\bu$ by
\begin{equation}
\bepsilon = \frac{1}{2}(\bdel\bu + (\bdel\bu)^{\top}).
\end{equation}
The stress tensor $\boldsymbol{\sigma}(\boldsymbol{x},t)$ is then a linear combination of the components of the strain tensor (Hooke's law):
\begin{equation}
\bsigma=\bc:\bepsilon , \label{eq:Hooke}
\end{equation}
where the colon denotes a double tensor contraction operation.
The elastic properties of the medium are described by the fourth-order elastic tensor $\boldsymbol{c}(\boldsymbol{x})$,
which can have up to $21$ independent coefficients.
In this article for simplicity we will consider isotropic media and put aside the more
general anisotropic relationship \eqref{eq:Hooke}. However anisotropy can readily be implemented (see for instance \citet{KoBaTr00b}).
In elastic isotropic media the stress tensor field reads
\begin{equation}
\boldsymbol{\sigma}=\lambda\Tr(\boldsymbol{\epsilon})\boldsymbol{I}+2\mu\boldsymbol{\epsilon} \, ,
\end{equation}
where $\boldsymbol{I}$ is the $3\times3$ identity tensor and $\Tr(\boldsymbol{\epsilon})$ is the trace of the strain tensor.
The Lam\'e parameters of the medium are denoted by $\lambda(\boldsymbol{x})$ and $\mu(\boldsymbol{x}).$
They are related to the pressure wave speed $c_p$, shear wave speed $c_s$ and density $\rho$ by the expressions
$\mu = \rho \, c_s^2$ and $\lambda = \rho \, c_p^2 - 2 \rho \, c_s^2$.

Likewise we will avoid the numerical complications associated with attenuation (viscoelasticity), however attenuation is
implemented in our code\cite{KoTr99,CrKo12} and we will use it in the numerical validation tests and examples of Section~\ref{sectionnumericalvalidation}.
Let us just recall that in an anelastic medium the stress $\boldsymbol{\sigma}(\boldsymbol{x},t)$ at time $t$
is determined by the entire strain history $\boldsymbol{\epsilon}(\boldsymbol{x},t)$, and Hooke's law becomes\cite{AkRi80,DaTr98}
\begin{equation}
\boldsymbol{\sigma}(\boldsymbol{x},t)=\int_{-\infty}^{t}\frac{\partial\boldsymbol{c}}{\partial t}(t-t'):\boldsymbol{\epsilon}(\boldsymbol{x},t')\:\mathrm{dt}'.
\end{equation}

As in the fluid part, one obtains the weak form by dotting the wave equation \eqref{eq:momentum} with a test function $\boldsymbol{w}$
and integrating by parts over the model volume $\Omega$. The only difference with the fluid formulation
is that the test function is vectorial $\boldsymbol{x}\mapsto\boldsymbol{w}(\boldsymbol{x})=(w_r(\boldsymbol{x}),w_z(\boldsymbol{x}))$. One obtains,
\begin{equation}
\int_{\Omega_s}\boldsymbol{w}\cdot\rho\ddot{\boldsymbol{u}}\:\mathrm{d^{2}\boldsymbol{x}} =
    -\int_{\Omega_s}\boldsymbol{\nabla w}:\boldsymbol{\sigma}\:\mathrm{d^{2}\boldsymbol{x}}
    +\int_{\Omega_{f-s}}\boldsymbol{w}\cdot(\boldsymbol{\sigma}\cdot\boldsymbol{n})\:\mathrm{d\Gamma} , \label{eq:weakForm3Delast}
\end{equation}
where
\begin{eqnarray}
\boldsymbol{\nabla w}:\boldsymbol{\sigma}                   & = & \sigma_{rr}\dfrac{\partial w_{r}}{\partial r}+\sigma_{zz}\dfrac{\partial w_{z}}{\partial z}
+ \sigma_{zr}\left(\dfrac{\partial w_{r}}{\partial z}+\dfrac{\partial w_{z}}{\partial r}\right)+\sigma_{\theta\theta}\dfrac{w_{r}}{r} , \\
\boldsymbol{w}\cdot(\boldsymbol{\sigma}\cdot\boldsymbol{n}) & = & (\sigma_{rr}n_r+\sigma_{rz}n_z)w_r+(\sigma_{zr}n_r+\sigma_{zz}n_z)w_z.
\end{eqnarray}
The important difference with a 2D planar model is the presence of the ``hoop stress'' term, $\sigma_{\theta\theta}$ - see Figure \ref{fig:cyl}.
Here again the traction-free integral along the free surface $\partial\Omega$ has vanished naturally.
For simplicity we have supposed that there is no source in the elastic part of the model.
It has to be noted that the terms ${u_r}/{r}$ and ${w_r}/{r}$ appear and are problematic in axial elements
for the GLJ point that is located exactly on the axis, i.e. at $r=0$.
However, on the axis, by symmetry considerations $u_r=0$ and, by definition of the test functions in axisymmetric domains\cite{BeDaMaAz99}, $w_r=0$.
Thus L'H\^opital's rule can be applied, as in the fluid case. For any $g$, which can represent $u_r$ or $w_r$ in practice, we define
\begin{equation}
\dfrac{g^{\overline{\sigma}\nu}}{r^{\overline{\sigma}\nu}}\equiv\begin{cases}
\dfrac{g^{\overline{\sigma}\nu}}{r^{\overline{\sigma}\nu}}, & \sigma \neq 0, \\
\left(\left.\dfrac{\partial r}{\partial\xi}\right|^{0\nu}\right)^{-1}
  {\displaystyle \sum_{\alpha,\beta=0}^{N}g^{\overline{\alpha}\beta}\overline{\ell}'_{\alpha}(\overline{\xi}_{0})\ell_{\beta}(\eta_{\nu})}, & \sigma=0.
\end{cases}
\end{equation}
Following similar steps as for the fluid part of the model, after assembling the system one obtains a similar global system of equations,
\begin{equation}
\left\{ \begin{array}{ccc}
{\displaystyle \ddot{\boldsymbol{U}}_{r}} & = & \left({\boldsymbol{M}^{g}}\right)^{-1}\boldsymbol{F}_{r,int}^{g}(t), \\
{\displaystyle {\displaystyle \ddot{\boldsymbol{U}}_{z}}} & = &
\left({\boldsymbol{M}^{g}}\right)^{-1}\boldsymbol{F}_{z,int}^{g}(t),
\end{array}\right. \label{eq:globalSystem2}
\end{equation}
where the $\boldsymbol{U}_i$ gather all the values of the (unknown) displacement and the $\boldsymbol{F}_{i,int}^g(t)$
gather all the interior forces at every GLL/GLJ point.

\section{Numerical validations}
\label{sectionnumericalvalidation}

We have tested the accuracy and efficiency of the technique and benchmarked it against reference solutions calculated with the wavenumber integration
software package OASES\cite{ScJe85,COA2011} version 3.1 as well as with the commercial finite-element code COMSOL.
The two examples include fluid-solid coupling, attenuation and PML absorbing layers;
they are extracted from \cite{JeNiZaCoSi07} and described in Figure \ref{fig:SettingSimulations}. Our code computes
time-domain signals but the results are shown both in time and frequency domains as an illustration.
For information, these examples ran in a few seconds (2~minutes maximum depending on the configuration chosen)
on 12 processor cores of an Intel Xeon$\textsuperscript{\textregistered}$ E5-2630 multicore PC.

\subsection{Flat-bottom benchmark case}
\label{subsection:benchmark}

\begin{figure}
\centerline{\includegraphics[width=0.495\linewidth]{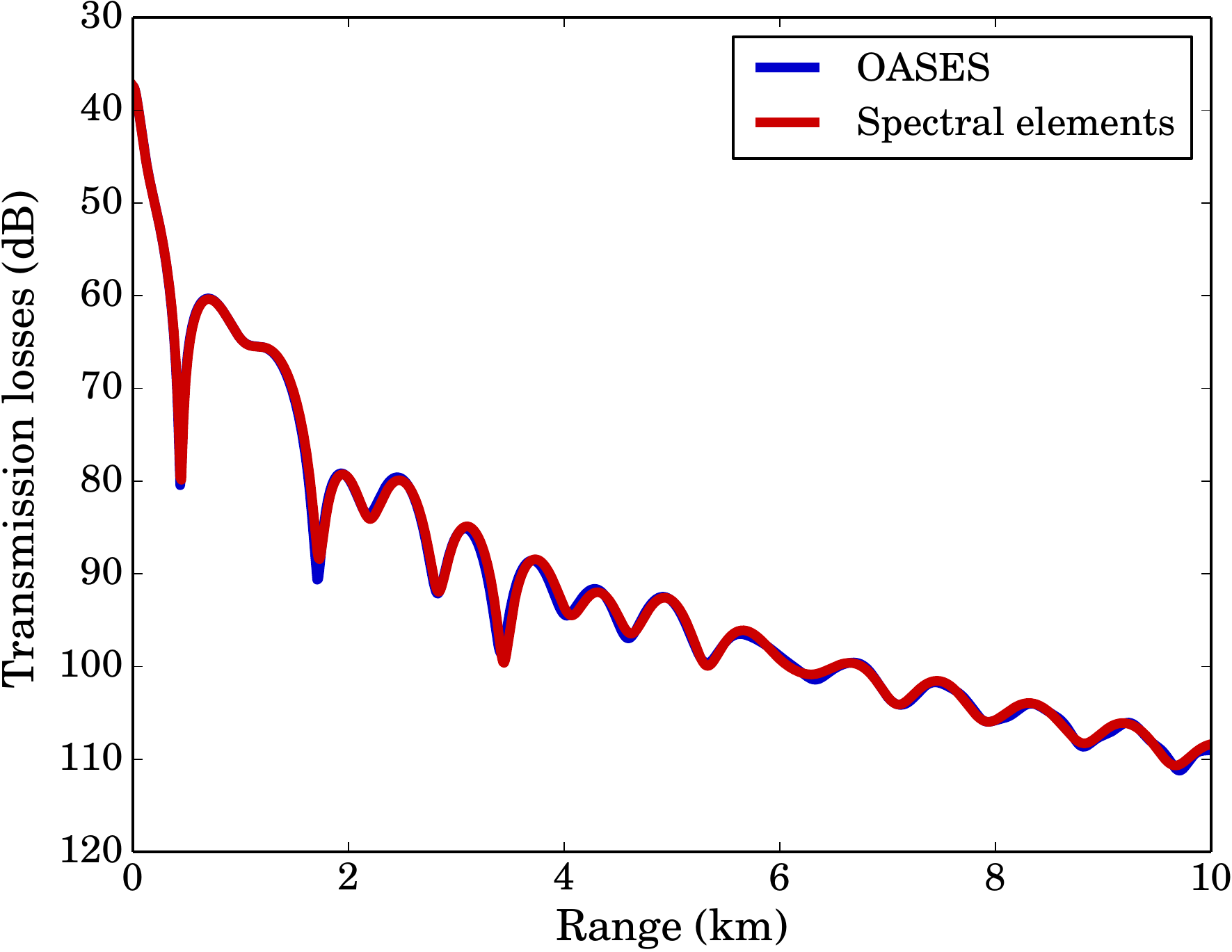}\hfill\includegraphics[width=0.495\linewidth]{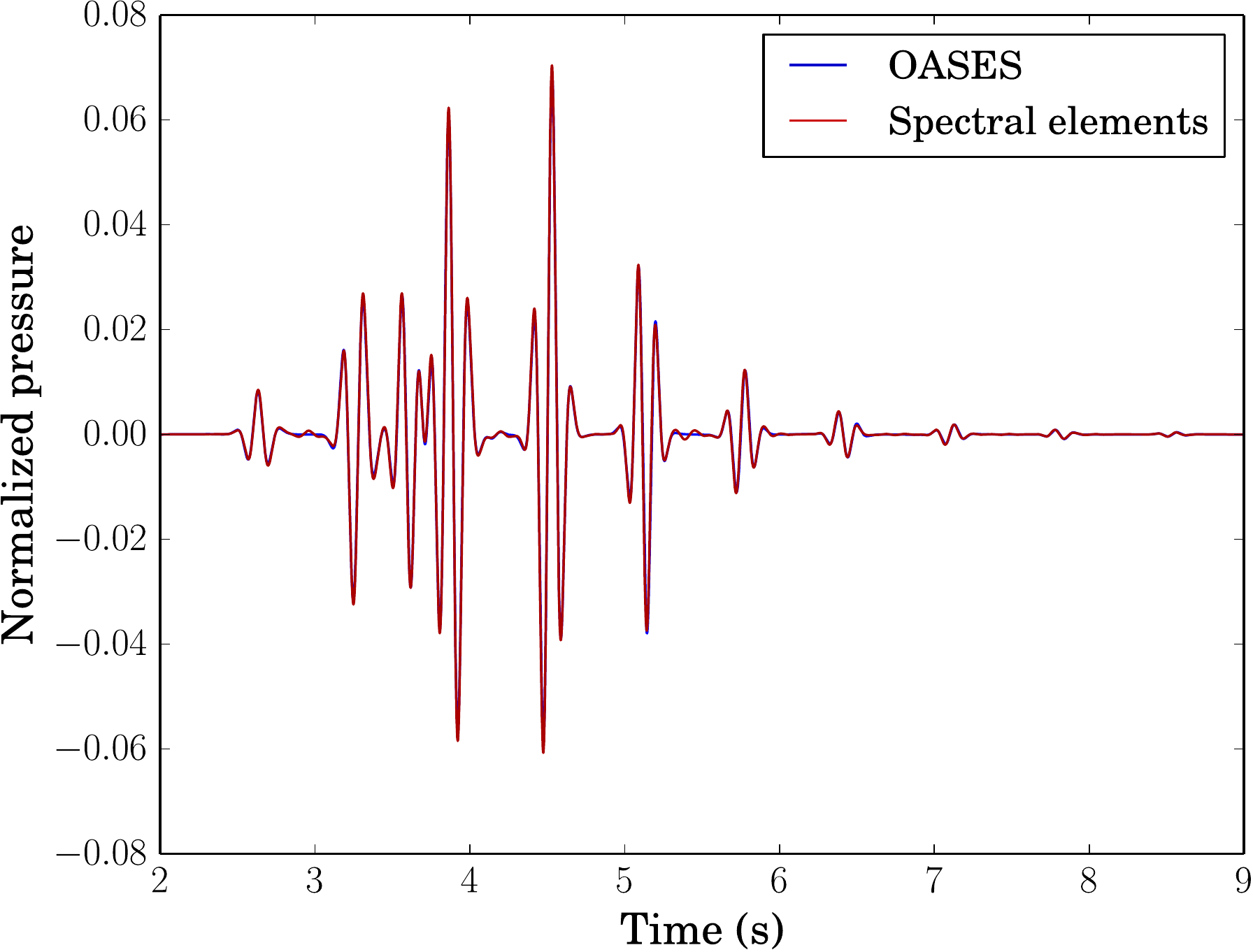}}
\caption{Comparison between the wavenumber integration code OASES and the spectral-element method for the flat-bottom benchmark case.
Left: Transmission losses in dB at 5 Hz and $z=-30$ m as a function of range in km.
Right: Normalized pressure (unit pressure at 1~m distance from the source) as a function of time at 5 km and $z=-30$ m.}
\label{fig:benchmark}
\end{figure}

Let us first compare the results of our spectral-element technique with those from the OASES wavenumber integration code for a flat-bottom benchmark case.
We consider an axisymmetric half-space composed of a semi-infinite homogeneous viscoelastic medium lying 600 m below a homogeneous sea layer.
The properties of the homogeneous viscoelastic part are given by $\rho=2000$ kg.m$^{-3}$ for the density,
$c_p=2400$ m.s$^{-1}$ for the pressure wave speed,
$c_s=1200$ m.s$^{-1}$ for the shear wave speed and $\alpha_p=0.2$ dB/$\lambda_P^{-1}$,
$\alpha_s=0.2$ dB/$\lambda_S^{-1}$ the corresponding attenuation coefficients.
In the acoustic domain we set the density to $\rho=1000$ kg.m$^{-3}$ and the pressure wave speed to $c_p=1500$ m.s$^{-1}$.
We set $z= 0$ m at the fluid surface.
The pressure source is located on the symmetry axis in the acoustic part of the medium, in $(r_s,z_s)$ = (0,-100 m);
its source time function is a Ricker (i.e. the second derivative of a Gaussian) wavelet with a dominant frequency $f_0 = 5$ Hz
and a time shift $t_0 = 1.2/f_0$ in order to ensure null initial conditions.
The wavefield is computed up to a range of 12 km and down to depth 1800 m,
the energy coming out of this box being absorbed by PMLs. The mesh is composed of $170 \times 60$ spectral elements whose polynomial degree is $N=4$.
The total number of unique GLL/GLJ points in the mesh is $(170N+1)\times(60N+1) = 164,121$.
The minimum number of grid points per shear wavelength in the solid is $5.2$
and the minimum number of points per pressure wavelength in the fluid is $6.8$.
We select a time step $\Delta t = 1.56$ ms and simulate a total of 6000 time steps, i.e. 9.36~s.
The pressure is recorded at $z=-30$ m by 1000 receivers uniformly distributed between $r=0$ m and $r=10$ km.
The pressure time series at 5 km and $z=-30$ m, normalized to unit pressure at a distance of 1~m from the source,
computed with OASES and with the spectral-element method are shown in Figure~\ref{fig:benchmark} (right)
and transmission losses in dB at 5 Hz and $z=-30$ m are shown in Figure~\ref{fig:benchmark} (left).
The transmission losses for the spectral-element method have been obtained from  the time signals by
Fourier transform. Very good agreement is found between the two codes in both the time and frequency domains,
thus validating our technique for this benchmark case.

\subsection{Validation, including for backscattering}
\label{subsection:slope}

\begin{figure}
\centerline{\includegraphics[width=0.8\linewidth]{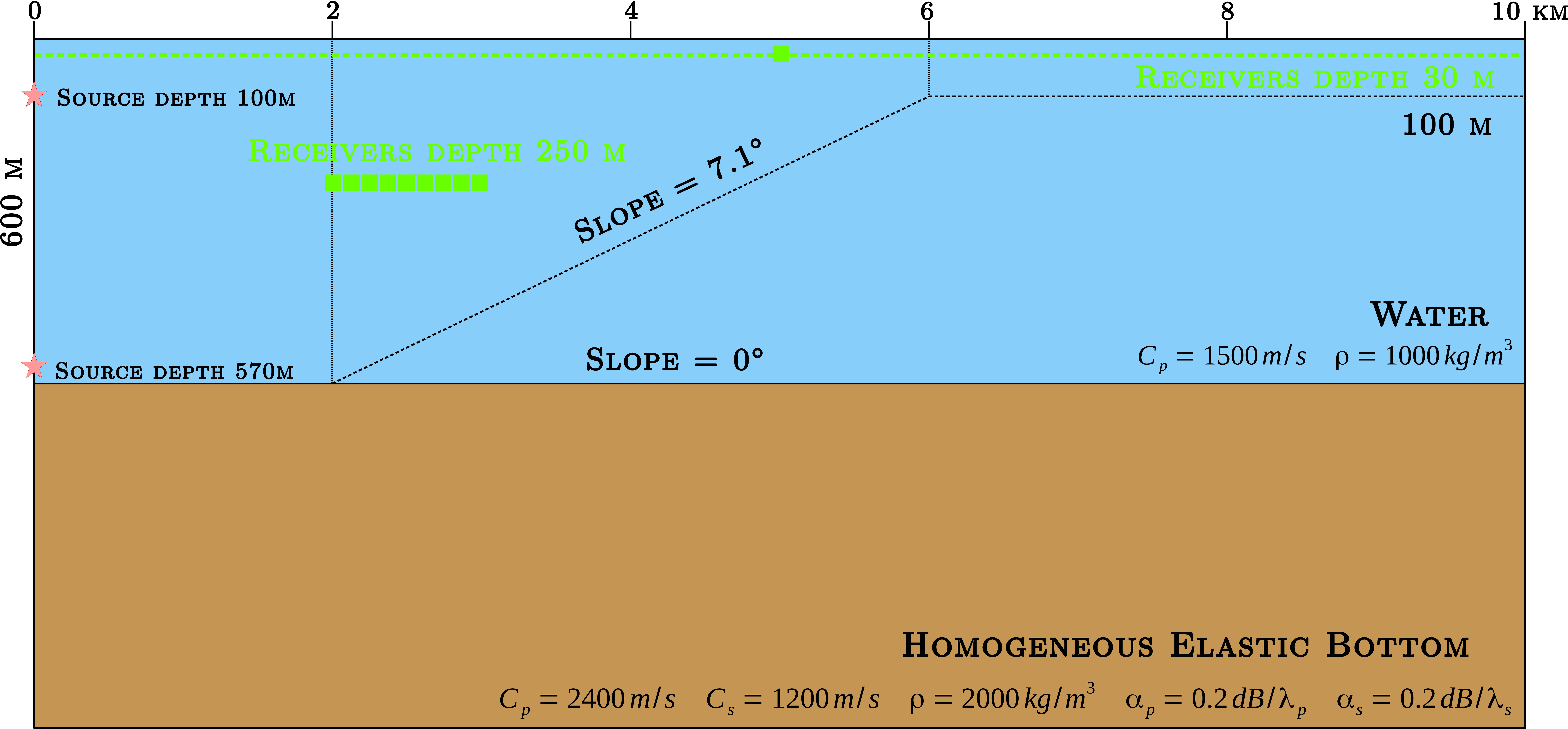}}
\caption{\label{fig:SettingSimulations} Setting of the validation simulations. The vertical and horizontal scales are different.
The properties of the two homogeneous media are given in the figure.
Slope = 0$^{\circ}$ corresponds to the flat bottom benchmark case of Section \ref{subsection:benchmark},
in which the water depth is constant and equal to 600 m.
In this case the source depth is 100 m and the receiver depth is 30 m.
The receiver that records the signal of Figure \ref{fig:benchmark} (right) is shown as a square at range 5 km and depth 30 m.
For the upslope case presented in Section \ref{subsection:slope} the model has a constant water depth (600 m)
up to 2 km followed by a constant bottom slope of 7.1$^{\circ}$ up to 6 km. The depth remains constant and equal to 100 m hereafter.
In this case the source depth is 570 m and the receiver depth is 30 m.
The horizontal antenna that records the pressure time series of Figure \ref{fig:comsol} (right) is shown as squares at
depth 250 m and at ranges 2--3 km.
The horizontal size of the computational domain is 12 km, but it includes an absorbing layer (PML) on the right side, which we purposely do not represent
because the wavefield in it is by definition non physical; The figures only show the physically-meaningful region, i.e. a domain of size 10 km.}
\end{figure}

\begin{figure}
\centerline{\includegraphics[width=0.495\linewidth]{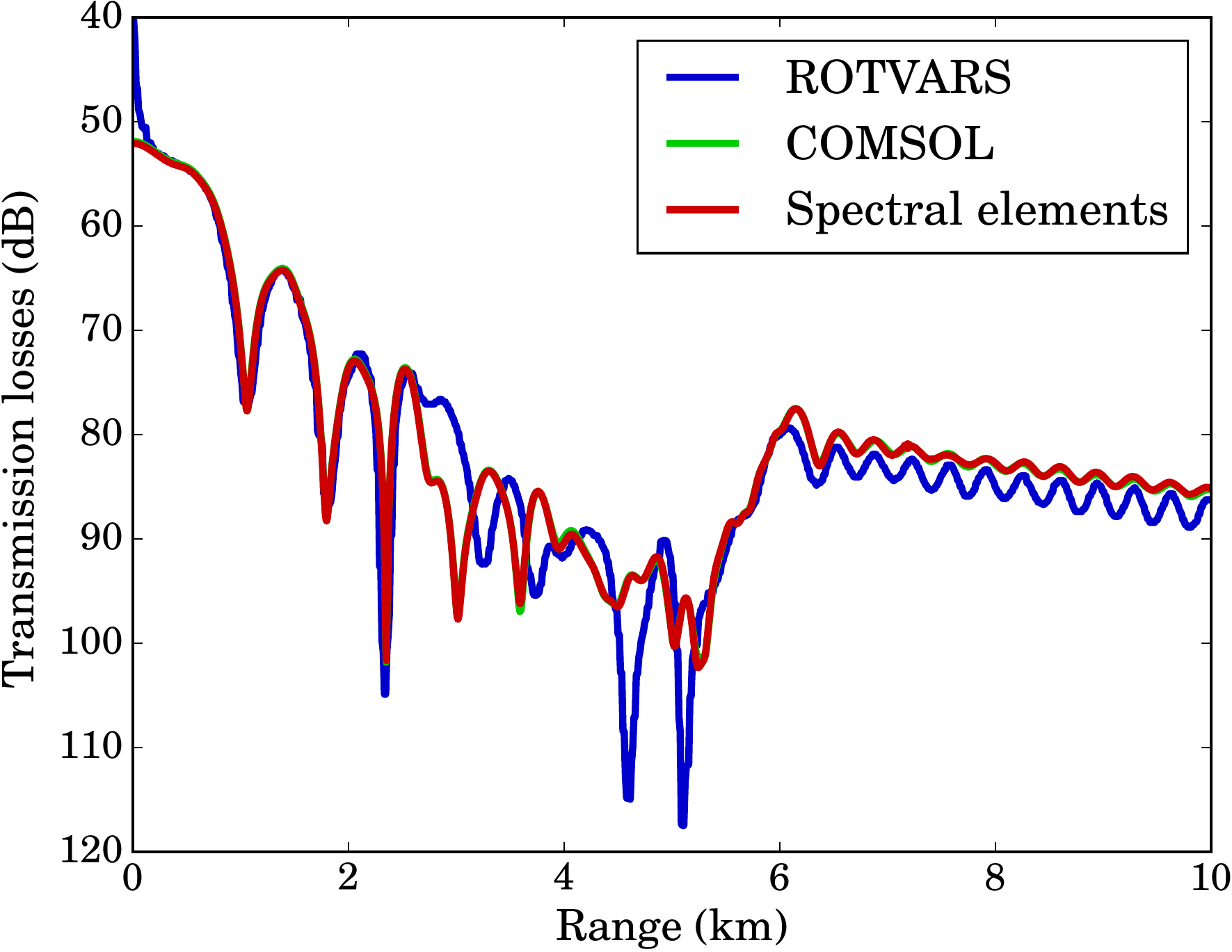}\hfill\includegraphics[width=0.495\linewidth]{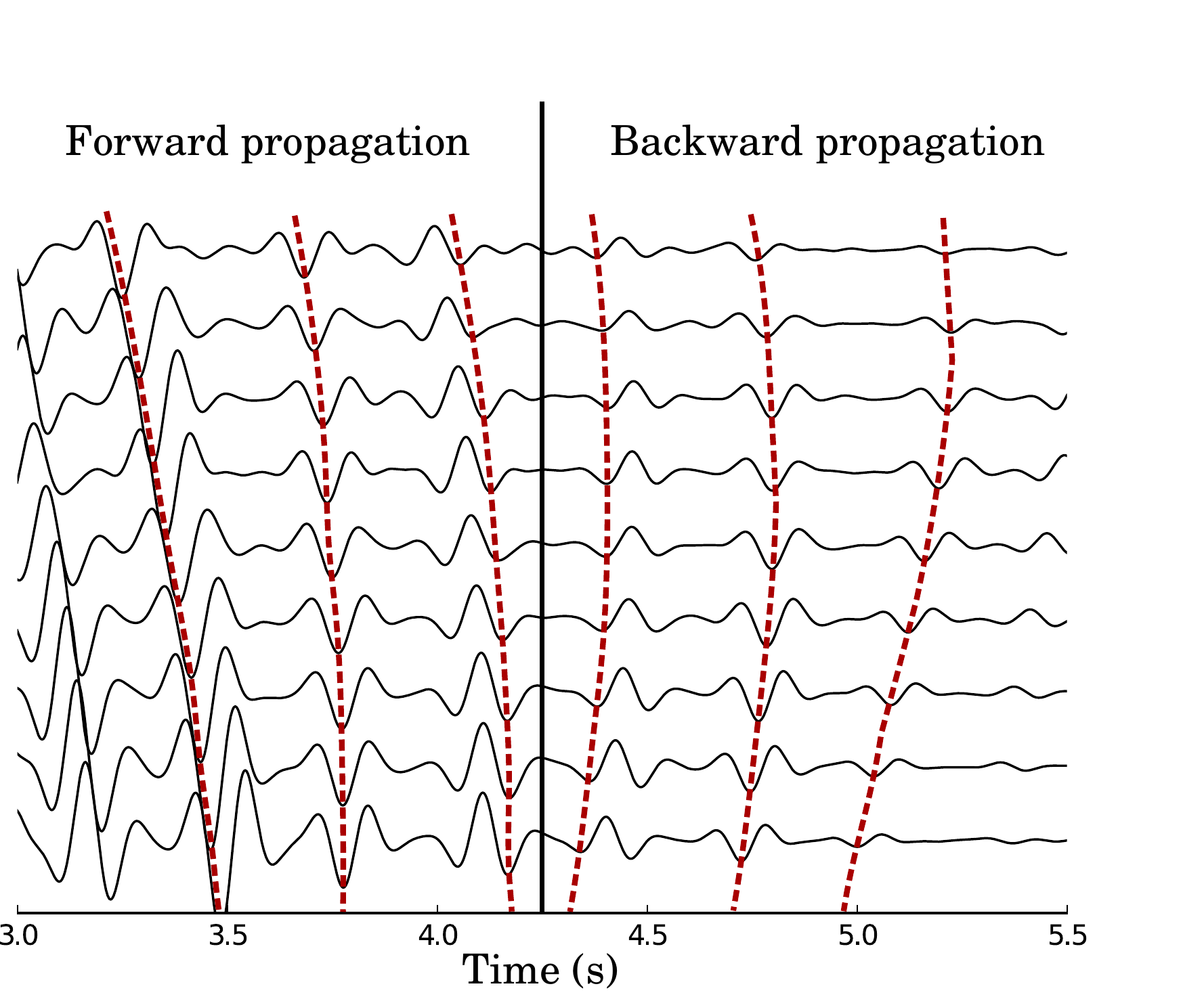}}
\caption{Left: Transmission losses in dB as a function of range in km.
The first curve has been calculated with ROTVARS, the second with COMSOL (both taken and adapted from \citet{JeNiZaCoSi07}, Fig. 7),
and the third with the spectral-element method. Note that the COMSOL and spectral-element results are almost perfectly superimposed.
Right: Pressure time series recorded by the receivers of the horizontal antenna ($z= -250$ m) shown in Figure \ref{fig:SettingSimulations}.
The upper trace is the pressure recorded by the receiver located at $r=2$ km, the lower trace is the pressure recorded by the receiver located at $r=3$ km.
In the left part of the picture ($t<4.25$~s) the noticeable arrivals are recorded first at the leftmost receivers, thus corresponding to
forward energy propagation. In the right part of the picture ($t>4.25$~s) the noticeable arrivals are recorded first at the rightmost receivers,
thus corresponding to significant backward energy propagation (which cannot be computed by parabolic equation techniques for instance).}
\label{fig:comsol}
\end{figure}

In this section we examine the results of our spectral-element method for the upslope case (slope $=$ 7.1$^{\circ}$) described in
Figure \ref{fig:SettingSimulations}. These results are compared to those from the commercial finite-element code COMSOL and
to a parabolic equation method with coordinate rotation (ROTVARS\cite{OuSiCoWe06}), specifically designed to handle variable slopes in bathymetry.
Both reference results were computed by \citet{JeNiZaCoSi07} and are reproduced here.
The properties of the media are the same as those used in the previous section.
The source depth is 570 m in order to excite a significant interface wave of Stoneley-Scholte type,
the shear wave speed being lower than the speed in water.
The source is still a pressure acoustic Ricker pulse with a dominant frequency $f_0 = 5$ Hz and a time shift $t_0 = 1.2/f_0$.
The wavefield is computed up to a range of 15 km and down to depth 3000 m,
the energy coming out of this box again being absorbed by PMLs. The mesh is composed of 9039 spectral elements whose polynomial degree is $N=4$
and the total number of unique GLL/GLJ grid points is 169,064. The number of grid points per shear wavelength in the solid is around 5.3
while the minimum number of grid points per pressure wavelength in the fluid is around 5.5. The time step chosen is $\Delta t = 1.62$ ms
and the total number of time steps is $8000$ in order to compute 12.96 seconds of simulation.
The pressure is recorded at $z=-30$ m by 1000 receivers uniformly distributed between $r= 0$ m and $r=10$ km
and by a horizontal antenna containing nine receivers (the squares at depth 250 m in Figure \ref{fig:SettingSimulations}).
The transmission losses in dB at 5 Hz and $z=-30$ m computed with COMSOL, ROTVARS
and with the spectral-element method are shown in Figure \ref{fig:comsol} (left).
An almost perfect fit is found between COMSOL and the spectral-element method,
even if the first is implemented in the frequency domain while the second is in the time domain.
All complex physical phenomena, including Stoneley-Scholte waves, are correctly modeled.
However the parabolic code ROTVARS does not line up because the slope bottom causes non negligible back-propagating waves that are not taken into account
by parabolic equation methods. This is illustrated in Figure \ref{fig:comsol} (right) showing the pressure time series recorded by the horizontal antenna.
This explains the differences already pointed out in \citet{JeNiZaCoSi07} and
illustrates the advantage of going beyond the parabolic solution for these kinds of problems.

\section{Example of a more realistic application to seamounts}

\begin{figure}
\centerline{\includegraphics[width=1\linewidth]{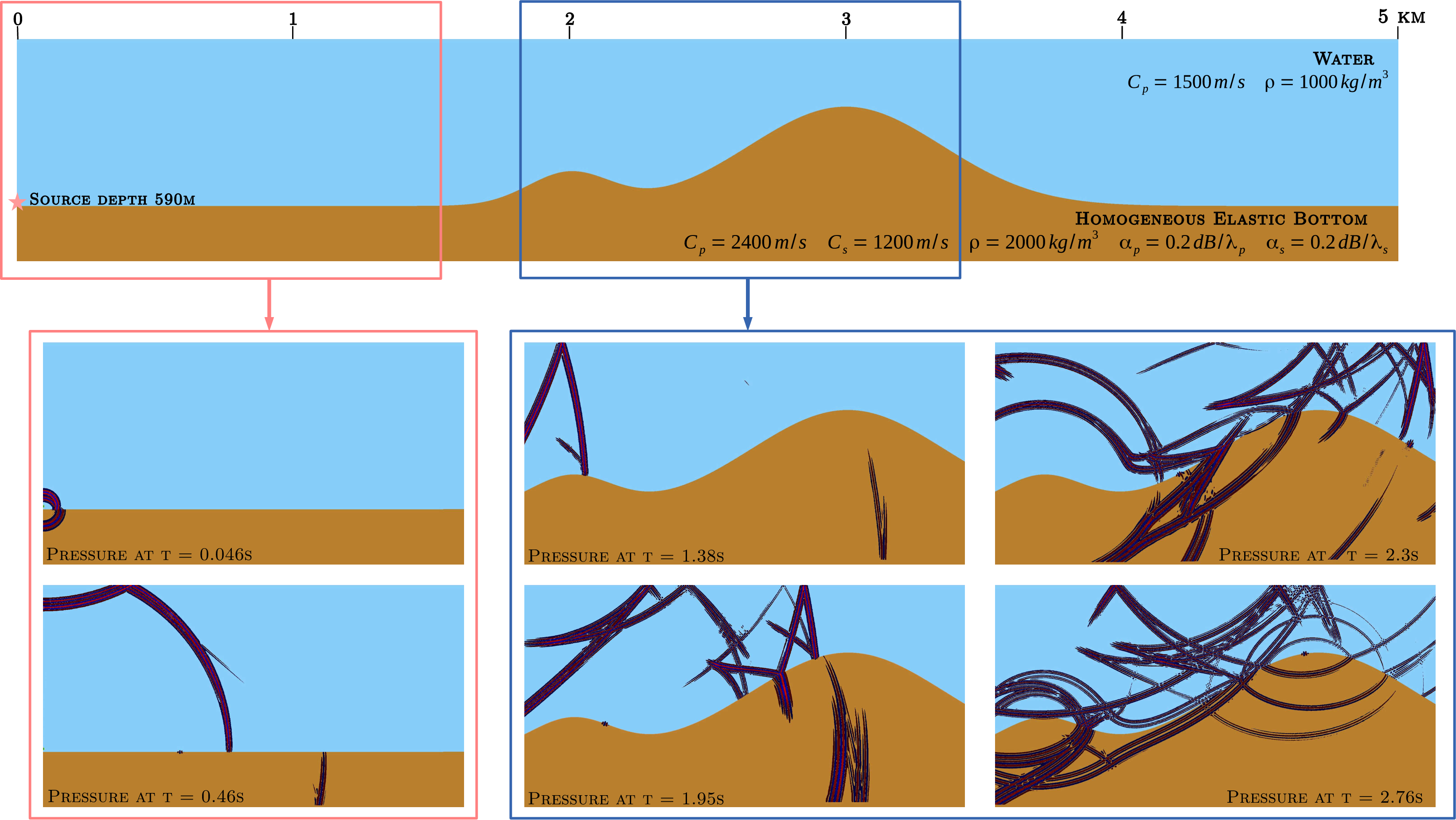}}
\caption{\label{fig:snapshots} Wavefield snapshots for an explosion with dominant frequency 100 Hz. The vertical and horizontal scales are
different.
The properties of the two homogeneous media are given in the upper figure. On the axis the water depth is 600 m,
and the source is located right above the ocean bottom, at a depth of 590 m.}
\end{figure}

In this section we present an example of a more realistic application to illustrate the possibilities offered by full-wave simulations.
The objective is to observe the wavefield behavior at long range created by an explosive source in an ocean with ocean-floor topography.
The model is described in Figure \ref{fig:snapshots}.
Performing a 2D axisymmetric simulation is very interesting for such cases with moderate azimuthal aperture\cite{Spi10} because,
since the source emits far from the seamounts, later effects are negligible and thus the results will be almost the same
as those of a fully 3D calculation, whose cost would still be prohibitive
even on current large supercomputers due to the high frequencies and large propagation distances involved (and will remain so for at least a decade or so,
as shown by the extrapolation of supercomputer power evolution that can be made for instance based on the data available in the Top500 database\cite{Top500}).
We keep the media properties identical to those of the previous section.
We again set $z= 0$ m at the fluid surface.
The pressure Ricker pulse source is located in $(r_s,z_s)$ = (0,-590 m) and has a dominant frequency $f_0 = 100$ Hz and a time shift $t_0 = 1.2/f_0$.
The wavefield is computed up to a range of 5 km and down to depth 1000 m,
the energy coming out of this box again being absorbed by PMLs. The mesh comprises $1600 \times 360$ spectral elements of polynomial degree $N=4$,
leading to a total number of unique GLL/GLJ points in the mesh of $(1600N+1)\times(360N+1) = 9,223,841$.
The minimum number of points per shear wavelength in the solid is 5.7 and
the minimum number of points per pressure wavelength in the fluid is 6.5. We use a time step $\Delta t = 0.046$ ms
and compute a total of 90,000 time steps, i.e. 4.14~s.
Here for simplicity we have chosen homogeneous media but media with complicated variations of their material properties can easily be accommodated\cite{KoTr99}.

Figure~\ref{fig:snapshots} illustrates the complexity of the wavefield structure obtained for this configuration. We consider two spatial windows, one covering a region
close to the source in which the sea floor is flat and another one that includes both seamounts. For each region we present snapshots of the wavefield at different times.
The first spatial window extends from the origin up to a range of 1.5 km. It shows the beginning of the propagation of the wavefield in a section of the
waveguide that is flat. The snapshot at time $t=0.046$~s shows the direct and reflected wavefronts very close to one another
because of the proximity of the source to the water-sediment interface, as well as a transmitted wavefront propagating into the sediment.
In the snapshot at time $t=0.46$~s the reflected and direct wavefronts have reached the water-air
interface, considered as a free surface in the simulation. A small wave packet that propagates along the interface, now separated from the two other wavefronts
and that corresponds to a surface wave of Stoneley-Scholte type, can be observed.
It can be identified because it propagates at a speed that is slower than all the wave speeds of the two media in contact ($c_{\mathrm{Stoneley}} = 1005$ m.s$^{-1}$).
The second spatial window covers a region starting at range 1.8 km and ending at range 3.4 km. The snapshot at time $t=1.38$~s exhibits a wavefront propagating in
the sediment (associated to the pressure wave) reaching the position of the second seamount while the wavefront propagating in the water
reaches the position of the first seamount. These wavefronts are the first arrivals.
At time $t=1.95$~s the influence of the two seamounts starts to affect wave propagation in the waveguide. A triplication
can be observed in the upper-left part of the snapshot owing to the shape of the first seamount. In the sediment, two wavefronts can be observed.
The first is associated to the first reflected wavefront from the water-air
interface transmitted into the sediment while the second comes from a transmission occurring at the first seamount because of the curvature
change. The same type of transmission into the sediment through the second seamount is beginning and can be seen half way to the top.
Some energy starts to penetrate into the sediment. This process continues and can be seen after the top of the second seamount at time $t=2.3$~s.
Two wavefronts above the first seamount correspond to the beginning of the significant backscattering that
occurs in this configuration. At time $t=2.76$~s multiple reflections generated between the water-air interface and the two seamounts interact and lead to a
complex wavefield structure. These multiple reflections are clearly seen at the top of the second seamount together
with the interface wave that was generated at the beginning of the simulation. Strong backscattering is observed.

For information, this example ran on 180 processor cores of a computing cluster in $\sim 30$ minutes
and could have run on a current classical high-end PC in less than a night.
It is worth mentioning that the spectral-element method exhibits almost perfect computing scaling i.e. almost perfect parallel efficiency on modern
parallel computers\cite{TrKoHjLiZhPeBoMcFrTrHu10,Kom11} when increasing the number of processors used
(mostly because its mass matrix is diagonal and thus no linear system needs to be inverted).

In terms of limitations of the approach let us mention again that the model being axisymmetric, calculations are made for seamounts whose 3D shape is annular.
This has to be kept in mind for axisymmetric simulations in particular when dealing with backscattered energy for objects located close to the symmetry axis
because for this geometrical reason backscattered energy gradually grows when coming closer to that axis and
is then reflected back into the model after having reached it; this comes from the fact that the radial component of displacement is zero by symmetry
on the symmetry axis and thus acts as a Dirichlet condition (perfectly reflecting condition) for the radial component of the field.
This is unphysical and constitutes a model error. A way of avoiding such fictitious `backscattering of the backscattering' by the axis
could maybe be to adopt a quasi-cylindrical formulation\cite{TaTaOkKe03,ToTaKa12}.
Another limitation of the approach is the type of sources that can be modeled if the source is located in the solid part of the model.
Since explosive sources have an axisymmetric shape they are perfectly modeled, whether they are located in the fluid or in the solid,
which allows for the study a large number of physical problems; however point force sources or moment tensor sources in solids\cite{TrKoHjLiZhPeBoMcFrTrHu10}
are limited to cases in which they have an axisymmetric radiation pattern.

\section{Conclusions and future work}

We have presented and validated a numerical method based on time-domain spectral elements for coupled fluid-solid full-wave propagation problems in an
axisymmetric setting, including in cases with significant backscattering,
which cannot be modeled e.g. based on the parabolic equation approximation.
Our calculations have included viscoelastic ocean bottoms, and we have used PML absorbing layers to efficiently absorb the outgoing wavefield.
An advantage of this method is its versatility and its relatively low computational cost compared to a truly 3D calculation while allowing simulations
with realistic 3D geometrical spreading.
An application to the calculation of the full wavefield at long range created by a high-frequency explosive source
in an ocean with sea-floor topography has also been presented.
Various future application domains are contemplated, among which we can cite more precise studies of seamounts
as well as the study of ocean T-wave dynamics\cite{JaGuGuMaRo13,FrCoOd15}.
Our SPECFEM spectral-element software package, which implements the spectral-element method presented above,
is collaborative and available as open source from the Computational Infrastructure for Geodynamics (CIG);
it includes all the tools necessary to reproduce the results presented in this article.

\section*{Acknowledgments}

We are grateful to Alexandre Fournier, Henrik Schmidt, Emmanuel Chaljub, Tarje Nissen-Meyer, Bruno Lombard, Chang-Hua Zhang and Zhinan Xie for fruitful discussions.
We thank Peter L. Nielsen and Mario Zampolli for providing their COMSOL and ROTVARS results from \citet{JeNiZaCoSi07},
and an anonymous reviewer for useful comments that improved the manuscript.
Part of this work was funded by the Simone and Cino del Duca / Institut de France / French Academy of Sciences Foundation under grant \#095164
and by the European `Mont-Blanc: European scalable and power efficient HPC platform based on low-power embedded technology' \#288777 project of call
FP7-ICT-2011-7. The Ph.D. grant of Alexis Bottero was awarded by ENS Cachan, France.
This work was also granted access to the French HPC resources of TGCC under allocation \#2015-gen7165
made by GENCI and of the Aix-Marseille Supercomputing Mesocenter under allocations \#14b013 and \#15b034.

%%%\bibliography{Biblio1}

\clearpage

\end{document}